\documentstyle[preprint,prl,aps]{revtex}
\begin{document}
\preprint{TNT 94-5}
\draft

\title{On the TAP approach to the Spherical p-spin SG Model}

\author{A. Crisanti}
\address{Dipartimento di Fisica, Universit\`a ``La Sapienza'', \\
         I-00185 Roma, Italy}

\author{H.-J. Sommers}
\address{Fachbereich Physik, Universit\"at-Gesamthochschule Essen,  \\
		 D-45141 Essen, Germany}

\date{June 7, 1994}

\maketitle

\begin{abstract}
In this letter we analyze the TAP approach to the spherical
$p$-spin spin glass model in zero external field.
The TAP free energy is derived by summing up all the relevant diagrams
for $N\to\infty$ of a diagrammatic expansion of the free energy.
We find that if the multiplicity of the TAP solutions is taken into
account, then there is a first order transition in the
order parameter at the critical temperature $T_{\rm c}$ higher than
that predicted by the replica solution $T_{\rm RSB}$, but in agreement
with the results of dynamics.
The transition is of ``geometrical'' nature since the new state
has larger free energy but occupies the largest volume in
phase space.
The transition predicted by the replica calculation is also of
``geometrical'' nature since it corresponds to the states with
smallest free energy with positive complexity.
\end{abstract}


The understanding of the low temperature phase of spin systems with
random couplings, namely spin glasses (SG), is still an open and
interesting problem. The main feature is the complex free energy
landscape made of many minima, separated by very high barriers, not
related by any symmetry one to another. This is responsible for the
non-trivial behavior of these systems, even at the mean field level
which is usually the first step towards the understanding of the phases
\cite{MPV87,FH91}.

Recently a simple mean field model \cite{CS92,CHS93} has been introduced
to investigate the static and dynamical properties of these systems.
This is an infinite range spherical SG
model with $p$-spin interactions. For any $p>2$ the model possess a
non-trivial low temperature and field phase. Within
the Parisi scheme of replica symmetry breaking the most general solution
for any temperature $T$ and field is obtained with only one step of
breaking (1RSB). In this paper we
shall consider the system without external field. In this case
the replica approach predicts a first order transition in the
order parameter at the critical temperature $T_{\rm RSB}$ where
the order parameter jumps discontinuously from zero (high temperature) to
a finite value. The free energy, nevertheless, remains continuous.
The study of the dynamics
yields a similar scenario but with a first order transition at an higher
critical temperature $T_{\rm c} > T_{\rm RSB}$, and a slightly different low
temperature phase. This  surprising result was first noted in Ref.
\onlinecite{KT87} in a soft-spin version of the model.
The reason why the two approaches led to two different results is that
in the replica approach the transition was obtained by the
requirement of largest replica free energy, while in dynamics it follows from
marginality. The two conditions are equivalent for the continuous
transition in a field, but not for the discontinuous one \cite{CS92,CHS93}.

In an attempt to understand this result Kurchan, Parisi and Virasoro
\cite{KPV93} proposed a TAP free energy for this model and showed
that in the absence of magnetic field the 1RSB solution is a
solution of the TAP equations. However, strangely enough, this solution
does not correspond to an extremum of the proposed TAP free energy.
Moreover at any temperature the replica symmetric solution leads to
a lower value of this free energy.
Therefore, it is not clear why there should be any transition.

We have derived the TAP free energy from a diagrammatic expansion of the
free energy by summing up all the relevant diagrams in the $N\to\infty$ limit.
In this letter we show that taking into account the degeneracy of the
TAP solutions, usually called ``complexity'', then
in the thermodynamic limit $N\to\infty$
this naturally leads to a transition in
agreement with the results of dynamics.
Moreover it gives the constraint under which
the TAP free energy for the 1RSB is minimal.
All details will be reported elsewhere.

The $p$-spin spherical SG model consists of $N$ continuous spins $\sigma_i$
interacting via quenched Gaussian couplings. The Hamiltonian is a $p$-body
interaction
\begin{equation}
  H(\sigma) = \frac{r}{2} \sum_{i=1}^N\, \sigma_i^2
     - \sum_{1\leq i_1<\cdots<i_p\leq N}\, J_{i_1,\ldots,i_p}\,
       \sigma_{i_1}\cdots\sigma_{i_p}
     - \sum_{i=1}^N h_i\sigma_i
\label{eq:ham}
\end{equation}
where we have included an external field $h_i$ and a parameter $r$ to
control the spin magnitude fluctuations.
The couplings are Gaussian variables with zero mean and average
$\overline{(J_{i_1,\ldots,i_p})^2} = p!/(2 N^{p-1})$. The scaling with $N$
ensures a well defined thermodynamic limit \cite{GM84}.
This formulation is slightly different from the
one given in Refs. \cite{CS92,KPV93}.
In the large $N$ limit the free energy per spin $f$ of the original spherical
model \cite{CS92,KPV93} and that of the model (\ref{eq:ham}), $\phi$, are
related by
\begin{equation}
  f(J,T,h) = \phi(r,J,T,h) - r/2
\label{eq:free}
\end{equation}
where $r$ is the value which makes the r.h.s. of (\ref{eq:free}) stationary.
This corresponds to impose the global constraint $\sum_{i=1}^N\sigma_i^2= N$
on the amplitude of the spins \cite{CS92,KPV93}.

Under general conditions the free energy $\phi$ can be derived from a
variational principle. To this end, we introduce the magnetization
$m_i = \langle\sigma_i\rangle$ and the connected spin-spin correlation
function
$G_{ij}= \langle\sigma_i\sigma_j\rangle
        - \langle\sigma_i\rangle\langle\sigma_j\rangle$
of (\ref{eq:ham}) for fixed couplings.
Then following Refs.\cite{CJT74,H91} the free energy $\phi$ can be
written as
\begin{equation}
 \beta N \phi(r,J,T,h) = H(m) + \frac{1}{2}{\rm Tr}\, \ln G^{-1}
                        + \frac{1}{2}{\rm Tr}\, {\cal D}^{-1}(m)\,G
                         - \Gamma_2(m,G) + \mbox{\rm const.}
\label{eq:phi}
\end{equation}
where
$T{\cal D}^{-1}(m) = \partial^2 H(\sigma)/ \partial\sigma_i\partial\sigma_j$
evaluated for $\sigma_i=m_i$.
The functional $\Gamma_2(m,G)$ is given by the sum of all two-particle
irreducible vacuum graphs in a theory with vertices determined by
(\ref{eq:ham}) and propagator set equal to $G_{ij}$.
The $m_i$ and $G_{ij}$ are evaluated at the stationary point of
the r.h.s. of (\ref{eq:phi}).

By taking into account only the diagrams which contribute to the averaged
free energy in the thermodynamic limit \cite{S78}, see Fig. \ref{fig:diag},
we obtain
\begin{eqnarray}
 \phi(q,g,E,r) = \frac{r}{2}(q+g) &-& \frac{1}{N}q^{1/2}\sum_i\, h_i\hat{m}_i
                + q^{p/2} E \nonumber \\
                &-& \frac{T}{2}\ln g
                - \frac{\beta}{4}\left[
                         (q+g)^p - q^p - pgq^{p-1}
                                 \right]
\label{eq:phi2}
\end{eqnarray}
where $Nq = \sum_i m_i^2$, $m_i = q^{1/2}\hat{m}_i$, $Ng = {\rm Tr}\, G$
and $E = - (1/Np!)\sum J_{i_1,\dots,i_p}\hat{m}_{i_1}\cdots\hat{m}_{i_p}$.
The details will be reported elsewhere.
In (\ref{eq:phi2}) we did not included the constant term which comes from
the normalization of the trace over the spin variables \cite{CS92} since
it does not change the stationary point.
In general $E$ is a random variable which
depends on both the  realization of couplings and on the orientation of the
vector  $\bbox{m}=(m_1,\ldots,m_N)$. Equation (\ref{eq:phi2}) is a
variational principle for the free energy since, for any value of $h$, $r$
and $T$, the $m_i$, or equivalently $q$ and $\hat{m}_i$, and $g$ are
determined by the stationary point of (\ref{eq:phi2}).

We shall now consider only the zero external field
case. From (\ref{eq:phi2}) we see that if $h_i=0$ all situations with the same
$E$ will lead to the same free energy. Consequently to have a defined problem
we consider $E$ as given and study the solution as function of $E$.
This corresponds to divide the phase space into classes according to the value
of $E$, and sum over all the states in one class. This is a statistics
by classification \cite{Ma}.

By eliminating $g$ and $r$ from the stationary point of equations
(\ref{eq:free}) and (\ref{eq:phi2})
\begin{eqnarray}
  g &=& 1 - q
\label{eq:sp_g} \\
  \beta r &=& \frac{1}{1-q} + \frac{\beta^2\,p}{2}\,(1-q^{p-1})
\label{eq:sp_r}
\end{eqnarray}
we are led to
the following variational principle for the free energy
per spin of the spherical $p$-spin SG model
\begin{equation}
 f(q,E,T)= q^{p/2} E - \frac{T}{2} \ln(1-q)
             - \frac{\beta}{4}\left[1+(p-1)q^p - pq^{p-1}\right]
\label{eq:tap}
\end{equation}
which, for any $T$ and $E$, has to be stationary with respect to $q$.
Equation (\ref{eq:sp_r}), for $\overline{r}(q) = \beta r - \beta^2\,p/2$,
is the equation of state first derived in \cite{CHS93} from the study of
dynamics. Here $r$ disappears being replaced by the free energy.

The functional $f(q,E,T)$ is related to the generating functional of
one-particle irreducible graphs, and hence the stationary point is
a minimum of $f$. Equation (\ref{eq:tap}) is the TAP free energy proposed
by Kurchan {\it et al} \cite{KPV93} and obtained by adding to the ``naive''
mean field free energy, the first two terms in (\ref{eq:tap}), the Onsager
reaction term, the last term in (\ref{eq:tap}), for the Ising $p$-spin SG
model \cite{rieger}.

The stationary point of equation (\ref{eq:tap}) gives, c.f.r. \cite{KPV93},
\begin{equation}
 (1-q)q^{p/2-1} = z T
\label{eq:sp}
\end{equation}
where
\begin{equation}
 z = \left[- E \pm \sqrt{ E^2 - E_{\rm c}^2} \right], \qquad
 E_{\rm c} = -\sqrt{2\,(p-1)/p}.
\label{eq:z}
\end{equation}

It is easy to understand that for any temperature $T$, and $z$ low enough,
there are two solutions of the saddle point equation (\ref{eq:sp}), one
corresponding to a maximum and one to a minimum. For
$z>z_{\rm c}=z(E_{\rm c})$ the stable
solutions leads to an unphysical $q$ decreasing with temperature. Therefore,
in (\ref{eq:z}) we take the `minus' sign. For $z<z_{\rm c}$, and $T$ low
enough, the stable solution is the largest one, $q\geq 1-2/p$.
The condition $z\leq z_{\rm c}$ is equivalent to the non negativity
of the relevant eigenvalue of the replica saddle point
\cite{CS92}. This confirms the assumption made in Ref. \cite{KPV93}
on the stability of the TAP solution. Here it comes naturally from
the analysis of the saddle point.

The results discussed so far are valid for any fixed $E\leq E_{\rm c}$. For
values larger than $E_{\rm c}$ there are no physical solutions.
The residual dependence of the free energy on $E$ follows from the
fact that we have summed only over all states within the class selected by
the given $E$. Consequently eq. (\ref{eq:tap}) represents the free energy
of that class. To have the full partition function we have to sum
$\exp(-\beta f)$, the partition function of the class $E$, over all
classes, including the degeneracy factor.
In the
thermodynamic limit the sum can be done by saddle point, so we have
\begin{equation}
 f_{J} = \min_{E_J}\, \left[ f(E_J,T) - \frac{T}{N}\ln{\cal N}(E_J) \right]
\label{eq:frej}
\end{equation}
where the subscript ``$J$'' denotes that all this has to be done for
fixed couplings. In other words the minimum has to be taken over all
allowed values of $E$ for the given realization of couplings.
In (\ref{eq:frej}) ${\cal N}(E_J)$ is the volume, or density of states,
of the class. In general ${\cal N}(E_J)$ is a random function which depends
on the disorder only through the value of $E_J$. This follows from the
fact that for fixed temperature $f$ depends only on the value of $E_J$.

By definition $f_J$ is a function of  the realization of disorder. However
the free energy is self-averaging. This means that for $N\to\infty$ the
overhalming majority of sample will give the same free energy
$\overline{f}(T)$, i.e,
\begin{eqnarray*}
  \mbox{\rm for}\ N\to\infty \qquad f_J(T) &=& \overline{f}(T) \qquad
  \mbox{\rm with probability $1$.}
\end{eqnarray*}
As a consequence $\overline{f}(T)$ can be obtained just averaging eq.
(\ref{eq:frej}) over disorder. Due to the selfaveraging of $E_J$ this means
to replace the second term by $\ln\overline{ {\cal N}(E)}$ and taking the
minimum over all allowed values of $E$.

We have calculated $\overline{ {\cal N}(E)}$ following the lines of
Refs. \cite{Bray,rieger}. The details will be reported elsewhere.
The explicit calculation reveals that
$g(E)= \ln\overline{{\cal N}(E)}/N$ for $N\to\infty$ is given by
\begin{equation}
  g(E) = \frac{1}{2}\biggl( \frac{2-p}{p} - \ln\frac{p\,z^2}{2}
                          + \frac{p-1}{2}\,z^2 - \frac{2}{p^2\,z^2}
                    \biggr), \qquad
             z \leq z_{\rm c}.
\label{eq:ge}
\end{equation}
This function is an increasing function of $E$ which takes its maximum
at the extremum $E=E_{\rm c}$ and is zero for $E=E_{\rm RSB} < E_{\rm c}$.
In figure \ref{fig:ge} we report the behavior of $g(E)$ as a function of
$E$ for $p=3$ where the corresponding values of $E_{\rm c}$ and
$E_{\rm RSB}$ are indicated.
For $E>E_{\rm c}$ there are no physical solutions, i.e. ${\cal N}(E)=0$.
For $E<E_{\rm RSB}$ the volume ${\cal N}(E)$ is exponentially small in $N$.
Consequently, in looking for the minimum in (\ref{eq:frej}) we have to
restrict ourselves to
values of $E$ in the range $E_{\rm RSB}\leq E\leq E_{\rm c}$.

Collecting all the results we have that for $N\to\infty$
there exists a critical temperature $T_{\rm c}$ below which
the thermodynamics of the spherical $p$-spin SG model is described by
the free energy
\begin{equation}
 \overline{f}(T) = f(q,E,T) - T\, g(E) - \frac{T}{2}\,[1+\ln(2\pi)]
\label{eq:thermo}
\end{equation}
where $q$ is given by eq. (\ref{eq:sp}) and $E$ is the value which for
the given temperature makes the r.h.s of (\ref{eq:thermo}) minimal.
The last term, not included before, comes from the normalization of
the trace over the spins and represents the entropy of the system
at infinite temperature \cite{CS92}.

The critical temperature $T_{\rm c}$ is the largest temperature where
$\overline{f}(T)$ is equal to the free energy of the replica symmetric
solution $q=0$ (high temperature solution), and is obtained for
$E=E_{\rm c}$. This corresponds to the critical temperature derived
from dynamics from marginality. Indeed for $E=E_{\rm c}$ we have
$z=z_{\rm c}$, i.e. the marginal condition \cite{CHS93}.

As the temperature is decreased the value of $E$ which minimizes
(\ref{eq:thermo}) decreases until it reaches the lower bound $E_{\rm RSB}$.
This happens at the critical temperature $T_{\rm RSB}$, the same as found
in the replica approach. From this point on the value of $E$ cannot be
decreased further since for lower values the number of solutions is
exponentially small in $N$. Therefore for $T<T_{\rm RSB}$ we have
$E=E_{\rm RSB}$.
We note that while for temperatures in the range
$T_{\rm RSB}\leq T\leq T_{\rm c}$ the free energy (\ref{eq:thermo}) is
numerically equal to that of the replica symmetric solution, for
$T<T_{\rm RSB}$ it is larger. Nevertheless it is the lowest free energy
among all the accessible states.

Similarly, in a dynamical calculation we have to restrict to the states
with an energy corresponding to the largest volume in the phase space
where the systems spends most of the time in its evolution.
In our case the volume is
proportional to $\exp(N\,g(E))$ which is maximal for $E=E_{\rm c}$, and all
other permitted states have exponentially small volume compared to this.
This means that the time to visit the other states is exponentially large
in the system size.
Therefore in the thermodynamic limit we have to restrict to
$E=E_{\rm c}$ and we get the free
energy (\ref{eq:thermo}) with $E$ replaced by $E_{\rm c}$ for which
\begin{equation}
  g(E_{\rm c}) = - \frac{1}{2}\left[ \ln(p-1) - 2\,\frac{p-2}{p} \right].
\label{eq:gec}
\end{equation}

The energy defined by the thermodynamic relation
${\cal E}= \partial \beta\overline{f}(T) / \partial\beta$ is not
affected by the complexity (\ref{eq:gec}) since it does not depend on
temperature.
Moreover, it turns out that the energy so defined is {\it equal} to
the energy derived from the dynamical calculation \cite{CHS93},
\begin{equation}
  {\cal E} = -\frac{\beta}{2}\,\left( 1 - q^p + m\,q^p \right)
\label{eq:ene}
\end{equation}
where the parameter $m$ following from the ``quasi fluctuation dissipation
theorem'' is obtained from the marginal condition: $m=(p-2)(1/q -1)$
with $q$ given by (\ref{eq:sp}) with $z=z_{\rm c}$.

We note that the free energies calculated in this paper
for different $E$ are all higher than free energies calculated in the
replica approach for different $m$. They coincide only for the 1RSB solution.
In Fig. \ref{fig:free} we report $\overline{f}(T)$ as a function of $T$ for
$p=15$ and different values of $E$. The free energies calculated in the
replica approach are all below the 1RSB free energy.
The free energy computed in this paper for the
marginal solution corresponds to the correct free energy of the dynamical
solution. It gives, in fact, the correct dynamical energy ${\cal E}$,
while the
corresponding quantity derived from the replica free energy for the
marginal $m$ gives a much lower energy.

In presence of an external field the scenario could be more complex since
it is not {\it a priori} clear if there exists a consistent free energy
corresponding to the dynamical state.

We conclude by noting that quite recently Marinari, Parisi and Ritort
found, in a different model, numerical evidence of the scenario
discussed in this paper \cite{Parisi}.

\acknowledgments
AC thanks the Sonderforschungsbereich 237 for financial support and the
Universit\"at-Gesamthochschule of Essen for kind hospitality, where part
of this work was done.

\begin{figure}
\caption{The two-particle irreducible diagrams which contribute in the
         limit $N\to\infty$ to $\Gamma_2(m,G)$. Each vertex has $p$ lines
         and gives a contribution $-\beta J_{i_1,\ldots,i_p}$.
         Each line joining two vertices gives a factor $G_{ij}$, while each
         ``dead-line'' gives a factor $m_i$.
        }
\label{fig:diag}
\end{figure}

\begin{figure}
\caption{$g(E)$ as a function of $E$ for $p=3$. The range of $E$ is
         restricted to $E\leq E_{\rm c}$. The value $E_{\rm RSB}$
         denotes the 1RSB solution.
        }
\label{fig:ge}
\end{figure}

\begin{figure}
\caption{$\overline{f}(T)$ eq. (\protect\ref{eq:thermo}) as a function of
         $T$ for $p=15$ and different values of $E$:
         (a) $E=E_{\rm c}$;
         (b) $E_{\rm RSB} < E < E_{\rm c}$;
         (c) $E=E_{\rm RSB}$;
         (d) the replica symmetric (high temperature) free energy.
        }
\label{fig:free}
\end{figure}

\end{document}